\documentclass[a4paper]{article}%
\usepackage{amsmath}
\usepackage{graphicx}%
\usepackage{amsfonts}%
\usepackage{amssymb}
\begin{document}

\author{D.L.Kovrizhin\\Russian Research Center, Kurchatov Institute,\\Kurchatov Square, 123182 Moscow, Russia}
\date{17 June 2001 }
\title{Exact form of the Bogoliubov excitations in one-dimensional nonlinear
Shr\"{o}dinger equation}

\begin{abstract}
In the paper we present the exact solutions of one-dimensional Nonlinear
Shr\"{o}dinger Equation. The solutions correspond to the Bogoliubov
excitations in Bose-gas with a local interaction. The obtained expression is
used for evaluating the transmission coefficient of the excitations across a
$\delta$-functional potential barrier.

\end{abstract}
\maketitle

\section{Introduction}

Bose-Einstein condensation has been observed in trapped dilute alkali-metal
atomic gases. The experimental developments have been impressive, and a number
of fundamental properties of these many-body systems can be investigated due
to the ideal possibilities for manipulation and monitoring of the condensates.
Theoretically they are well treated by the simplest approximate theories of
many-body physics, and they are therefore ideal for investigating properties
of simple many-body systems \cite{Pitaevskii}. These systems are well
described by the Gross-Pitaevskii equation \cite{Landau},\cite{cher} which
transforms to Nonlinear Shr\"{o}dinger Equation (NLS) in a local limit.
Nonlinear Shr\"{o}dinger Equation which is one of the fundamental equations of
nonlinear physics \cite{lamb} is widely used in many fields of modern science
from nonlinear optics to elementary particle physics. Localized solutions of
NLS which correspond to nonreflective potentials of scattering problem are
well-investigated. It is used to employ Bogoliubov's method of decomposition
of time-dependent wave function by wave functions of elementary excitations.
One can find for the last a whole system of second order differential
equations, which is known as a Bogoliubov-deGennes system. There are methods
of exact investigation of such equations\textbf{ }\cite{lamb}. In the paper it
is shown explicit form of exact solutions, which wasn't found in literature.
As an\textbf{ }example the obtained expression is used for evaluating the
transmission coefficient of the excitations across a $\delta$-functional
potential barrier. The result is interesting in connection with investigations
of Josephson-like effect in BEC.

\section{Bogoliubov-de Gennes system of equations}

We consider a zero-temperature Bose-Einstein condensate of atoms with mass $m$
and chemical potential $\mu$. The atoms interact by elastic collisions with
s-wave scattering length $a_{sc}$ and their low kinetic energies permit a
replacement of their short-range interaction by a contact term $V_{int}\left(
\mathbf{r}\right)  \rightarrow U_{0}\delta\left(  \mathbf{r}\right)
=4\pi\hbar^{2}a_{sc}\delta\left(  \mathbf{r}\right)  /M$, so that the single
particle wave functions obeys the Gross-Pitaevskii equation \cite{molmer}%
\begin{equation}
i\hbar\frac{\partial\Psi\left(  \mathbf{r},t\right)  }{\partial t}=\left[
-\frac{\hbar^{2}}{2m}\mathbf{\nabla}^{2}-\mu+U_{0}\left|  \Psi\left(
\mathbf{r},t\right)  \right|  ^{2}\right]  \Psi\left(  \mathbf{r},t\right)  .
\label{gpe}%
\end{equation}
Using proper units of measurement we will have in one dimension
\begin{equation}
i\frac{\partial\Psi(x,t)}{\partial t}=-\frac{1}{2}\frac{\partial^{2}\Psi
(x,t)}{\partial x^{2}}-\Psi(x,t)+\left|  \Psi(x,t)\right|  ^{2}\Psi(x,t).
\label{a4}%
\end{equation}
Equation (\ref{a4}) can be linearized using representation of $\Psi(x,t)$ as a
sum of time-independent wave function $\Psi_{0}(x)$ which describes the ground
state of the condensate and time-dependent wave function of the excitations
$\psi(x,t),$%
\begin{equation}
\Psi(x,t)=\Psi_{0}(x)+\psi(x,t). \label{a5}%
\end{equation}
Substitute (\ref{a5}) into (\ref{a4}) and assume $\psi(x,t)$ is a small value
in comparison with $\Psi_{0}(x)$ we will have the equation
\begin{equation}
-\frac{1}{2}\frac{d^{2}\Psi_{0}(x)}{dx^{2}}-\Psi_{0}(x)+\Psi_{0}^{3}\left(
x\right)  =0
\end{equation}
The exact well-known solution of this equation is the instanton
\cite{polyakov}%
\begin{equation}
\Psi_{0}(x)=\tanh x. \label{tan1}%
\end{equation}
Time-dependent wave function $\psi(x,t)$ is described by the equation%
\begin{equation}
i\frac{\partial\psi(x,t)}{\partial t}=-\frac{1}{2}\frac{\partial^{2}\psi
(x,t)}{\partial x^{2}}-\psi(x,t)+g\left(  x\right)  (2\psi(x,t)+\psi^{\ast
}(x,t)) \label{a6}%
\end{equation}
where $g\left(  x\right)  =\left|  \Psi_{0}(x)\right|  ^{2}=\tanh^{2}x$.

Let's find the solution of the equation (\ref{a6}), representing the wave
function of the excited state as a superposition of wave functions of the
elementary excitations,%
\begin{equation}
\psi\left(  x,t\right)  =u\left(  x\right)  e^{-i\varepsilon t}-v^{\ast
}\left(  x\right)  e^{i\varepsilon t}\label{psic}%
\end{equation}
Substituting (\ref{psic}) into (\ref{a6}), we will have Bogoliubov-de Genes
system of equations%
\begin{align}
\varepsilon u\left(  x\right)   &  =-\frac{1}{2}\frac{\partial^{2}u\left(
x\right)  }{\partial x^{2}}-u\left(  x\right)  +g\left(  x\right)  (2u\left(
x\right)  -v\left(  x\right)  )\label{bdgeq1}\\
-\varepsilon v\left(  x\right)   &  =-\frac{1}{2}\frac{\partial^{2}v\left(
x\right)  }{\partial x^{2}}-v\left(  x\right)  +g\left(  x\right)  (2v\left(
x\right)  -u\left(  x\right)  )\label{bdgeq2}%
\end{align}
We may obtain a simpler equation for the excitations, if we rewrite the
Bogoliubov-de Gennes equations,%
\begin{align}
\varepsilon T\left(  x\right)   &  =-\frac{1}{2}\frac{d^{2}S\left(  x\right)
}{dx^{2}}+\left(  g\left(  x\right)  -1\right)  S\left(  x\right)
\label{TT1}\\
\varepsilon S\left(  x\right)   &  =-\frac{1}{2}\frac{d^{2}T\left(  x\right)
}{dx^{2}}+\left(  3g\left(  x\right)  -1\right)  T\left(  x\right)
\label{SS1}%
\end{align}
where we have introduced \cite{molmer}%

\begin{equation}
S\left(  x\right)  =u\left(  x\right)  +v\left(  x\right)  ,\;T\left(
x\right)  =u\left(  x\right)  -v\left(  x\right)  .\label{STtrans}%
\end{equation}

The system (\ref{SS1},\ref{TT1}) can be transformed to the fourth order
differential equation%
\begin{equation}
\left(  -\frac{1}{2}\frac{d^{2}}{dx^{2}}+3\tanh^{2}x-1\right)  \left(
-\frac{1}{2}\frac{d^{2}}{dx^{2}}+\tanh^{2}x-1\right)  S\left(  x\right)
=\varepsilon^{2}S\left(  x\right)  \label{S4}%
\end{equation}
Let's find a solution of the equation (\ref{S4}) by Bargman's method
\cite{lamb} as a composition of plane wave and a some function, which is the
polynomial of $k$%
\begin{equation}
S\left(  x\right)  =F\left(  k,x\right)  e^{ikx}. \label{Sx}%
\end{equation}
The simplest nontrivial example has the form%
\begin{equation}
S\left(  x\right)  =\left(  \beta k+a\left(  x\right)  \right)  e^{ikx},
\label{Sx1}%
\end{equation}
where $a\left(  x\right)  -$ is the arbitrary sufficiently smooth function.
Substituting (\ref{Sx1}) into equation (\ref{S4}) and equating the components
at the same powers of $k,$ we will find the solution,%

\begin{equation}
S\left(  x\right)  =\left(  -\frac{1}{2}ik+\tanh x\right)  e^{ikx}.
\label{solS}%
\end{equation}
Where $k$ is one of the roots of equation%
\begin{equation}
\varepsilon=\frac{1}{2}k\sqrt{k^{2}+4}, \label{ek}%
\end{equation}
that determines the dispersion law of Bogoliubov's excitations. With small $k$
(\ref{ek}) gives the sound spectrum $\varepsilon=k$. The equation (\ref{S4})
is the fourth order linear differential equation and it must have four
linearly independent solutions. The equation (\ref{ek}) also has four
different solutions,%

\begin{align}
k_{1,2} &  =\pm\sqrt{2}\sqrt{-1+\sqrt{1+\varepsilon^{2}}},\label{kk}\\
k_{3,4} &  =\pm i\sqrt{2}\sqrt{1+\sqrt{1+\varepsilon^{2}}},\label{qq}%
\end{align}
that determines $k$ with given $\varepsilon.$ With $\varepsilon\ll1$ the
solutions (\ref{kk}) $k_{1,2}\approx$ $\pm\varepsilon$, but (\ref{qq}) are
pure imaginary $k_{3,4}\approx\pm2i.$ The general solution of (\ref{S4}) is a
linear superposition of solutions (\ref{solS}) with arbitrary coefficients
$C_{n}$%
\begin{equation}
S\left(  x\right)  =\sum_{n=1}^{4}C_{n}\left(  -\frac{1}{2}ik_{n}+\tanh
x\right)  e^{ik_{n}x}.\label{SF}%
\end{equation}
Substituting (\ref{SF}) into (\ref{TT1})\ we will have for $T\left(  x\right)
$%
\begin{equation}
T\left(  x\right)  =\sum_{n=1}^{4}C_{n}\frac{k_{n}}{2\varepsilon}\left(
k_{n}\left(  -\frac{1}{2}ik_{n}+\tanh x\right)  -i\operatorname{sech}%
^{2}x\right)  e^{ik_{n}x}.\label{solT}%
\end{equation}
Expressions (\ref{SF}) and (\ref{solT}) are total system of solutions for
(\ref{S4}). When the excitation moves in a Bose-condensate there is no
reflection from a kink because the wave function of the excitation is an
eigenfunction in a system with kink.

\section{Excitations scattering on a $\delta$ - functional potential barrier}

As an example let's use the obtained solutions to calculate the transmission
coefficient of excitations scattering on a $\delta-$functional barrier. The
problem is interesting in connection of the Josephson-like effect in BEC. It
was solved at small $k$ in \cite{kov}. The ground state of the equation,
corresponding to a Bose-gas with a local interaction in one dimension with an
external $\delta-$functional potential%
\begin{equation}
-\frac{1}{2}\frac{d^{2}\Psi}{dx^{2}}-\Psi+V_{0}\delta\left(  x\right)
\Psi+\left|  \Psi\right|  ^{2}\Psi=0 \label{eq81}%
\end{equation}
we will search in the form%
\begin{equation}
\Psi\left(  x\right)  =\tanh\left(  \left|  x\right|  +x_{0}\right)  .
\label{Psid}%
\end{equation}
The wave function must be continuous at $x=0$ but a derivative has a break,%
\[
\Psi^{^{\prime}}\left(  +0\right)  -\Psi^{^{\prime}}\left(  -0\right)
=2V_{0}\Psi\left(  0\right)  .
\]
It follows that%
\begin{equation}
1-\xi^{2}=V_{0}\xi,\qquad\xi=\tanh x_{0} \label{121}%
\end{equation}

The problem of excitations scattering on a local potential is described by the
solution of the equation (\ref{SF}), which at large negative values of
coordinate must transform to a superposition of \ incident\ and reflected
wave, but at large positive values of $x$ --- to transmitted wave. Let's write
the solutions at $x<0$ and $x>0$, satisfying necessary boundary conditions%

\begin{gather}
S\left(  x\right)  =e^{ikx}\left(  -\frac{1}{2}ik-\tanh\left(  \left|
x\right|  +x_{0}\right)  \right)  +Ge^{-ikx}\left(  \frac{1}{2}ik-\tanh\left(
\left|  x\right|  +x_{0}\right)  \right)  +\label{Sl0}\\
+Pe^{qx}\left(  -\frac{1}{2}q-\tanh\left(  \left|  x\right|  +x_{0}\right)
\right)  ,x<0\label{Sg0}\\
T\left(  x\right)  =\frac{k}{2\varepsilon}e^{ikx}\left(  k\left(  -\frac{1}%
{2}ik-\tanh\left(  \left|  x\right|  +x_{0}\right)  \right)  -i\left(
1-\tanh^{2}\left(  \left|  x\right|  +x_{0}\right)  \right)  \right)
+\label{Tl0}\\
+G\frac{k}{2\varepsilon}e^{-ikx}\left(  k\left(  \frac{1}{2}ik-\tanh\left(
\left|  x\right|  +x_{0}\right)  \right)  +i\left(  1-\tanh^{2}\left(  \left|
x\right|  +x_{0}\right)  \right)  \right)  -\label{Tg0}\\
-P\frac{q}{2\varepsilon}e^{qx}\left(  q\left(  -\frac{1}{2}q-\tanh\left(
\left|  x\right|  +x_{0}\right)  \right)  +\left(  1-\tanh^{2}\left(  \left|
x\right|  +x_{0}\right)  \right)  \right)  ,x<0\nonumber\\
S\left(  x\right)  =Fe^{ikx}\left(  -\frac{1}{2}ik+\tanh\left(  x+x_{0}%
\right)  \right)  +Qe^{-qx}\left(  \frac{1}{2}q+\tanh\left(  x+x_{0}\right)
\right)  ,x>0\nonumber\\
T\left(  x\right)  =F\frac{k}{2\varepsilon}e^{ikx}\left(  k\left(  -i\frac
{1}{2}k+\tanh\left(  x+x_{0}\right)  \right)  -i\left(  1-\tanh^{2}\left(
x+x_{0}\right)  \right)  \right)  +\nonumber\\
+\frac{q}{2\varepsilon}e^{-qx}\left(  -q\left(  \frac{1}{2}q+\tanh\left(
x+x_{0}\right)  \right)  +\left(  1-\tanh^{2}\left(  x+x_{0}\right)  \right)
\right)  ,x>0\nonumber
\end{gather}
where,%
\begin{align*}
k  &  =\sqrt{2}\sqrt{-1+\sqrt{1+\varepsilon^{2}}}\approx\varepsilon\\
q  &  =\sqrt{2}\sqrt{1+\sqrt{1+\varepsilon^{2}}}\approx2
\end{align*}
Using the\ lacing conditions of functions and its derivatives at zero and
assuming (\ref{121}) we will have the system of four linear equations with
four indeterminate. Solution of the system in approach $q=2$ gives the
following for the transmission and reflection coefficients%

\begin{align}
D  &  =\left|  F\right|  ^{2}=\xi^{2}\frac{\left(  4\xi^{3}+4\xi^{2}+2k^{2}%
\xi+4\xi+6k^{2}+4+k^{4}\right)  ^{2}}{\left(  k^{2}+4\right)  \left(  \xi
^{2}k^{2}+\allowbreak\left(  1+\xi^{2}\right)  ^{2}\right)  \left(
k^{4}+4k^{2}+4k^{2}\xi+4\xi^{2}+8\xi^{3}+4\xi^{4}\right)  }\label{DG}\\
R  &  =\left|  G\right|  ^{2}=k^{2}\frac{\left(  \xi^{2}-1\right)  ^{2}\left(
2\xi^{2}+2\xi+4+k^{2}\right)  ^{2}}{\left(  k^{2}+4\right)  \left(  \xi
^{2}k^{2}+\allowbreak\left(  1+\xi^{2}\right)  ^{2}\right)  \left(
k^{4}+4k^{2}+4k^{2}\xi+4\xi^{2}+8\xi^{3}+4\xi^{4}\right)  }\nonumber
\end{align}
The transmission coefficient at $k\gg1$ equals to $1$, and at $k\rightarrow0$
tends to $1$ as%
\begin{equation}
D=\frac{\xi^{2}}{\xi^{2}+k^{2}}, \label{Dk}%
\end{equation}
as in \cite{kov}.

Author thanks L.A.Maximov for his interest to the paper.

\end{document}